\documentclass{acm_proc_article-sp}
\usepackage{graphicx}
\usepackage{algorithmic}
\usepackage{algorithm}
\usepackage{ amsmath}
\usepackage[table]{xcolor}
\usepackage{multirow}
\long\def\comment#1{}

\newcommand{\denselist}{
     \setlength{\itemsep}{0pt}
     \setlength{\parsep}{1.5pt}
     \setlength{\topsep}{1.5pt}
     \setlength{\parskip}{2pt}
     \setlength{\partopsep}{0pt}
     \setlength{\labelwidth}{1em}
     \setlength{\labelsep}{0.5em} }

\newcommand{\bdesc}{\begin{description}\denselist}
\newcommand{\edesc}{\end{description}}

\newcommand{\blist}{\begin{itemize}\denselist}
\newcommand{\elist}{\end{itemize}}
\long\def\remove#1{}

\begin{document}
\title{Entropy-based Classification of `Retweeting' Activity on Twitter
\titlenote{The 4th SNA-KDD Workshop '11 ( SNA-KDD'11), August 21, 2011, San Diego CA USA . Copyright 2011 ACM 978-1-4503-0225-8...\$5.00.}
}

\numberofauthors{3}

\author{
\alignauthor Rumi Ghosh\\
       \affaddr{USC Information Sciences Institute}\\
       \affaddr{Marina del Rey, CA 90292}\\
       \email{\{rumig\}@isi.edu}
\alignauthor Tawan Surachawala\\
       \affaddr{USC Information Sciences Institute}\\
       \affaddr{Marina del Rey, CA 90292}\\
       \email{\{tawans\}@isi.edu}
\alignauthor Kristina Lerman\\
       \affaddr{USC Information Sciences Institute}\\
       \affaddr{Marina del Rey, CA 90292}\\
       \email{\{lerman\}@isi.edu}
}
\date{}
\maketitle
\begin{abstract}
Twitter is used for a variety of reasons, including information dissemination, marketing, political organizing and to spread propaganda, spamming, promotion, conversations, and so on. Characterizing these activities and categorizing associated user generated content is a challenging task. We present a  information-theoretic approach to classification of user activity on Twitter.  We focus on tweets that contain embedded URLs and study their  collective `retweeting' dynamics. We identify two features, time-interval and user entropy, which we use to classify retweeting activity.  We achieve good separation of different activities using just these two features and are able to categorize content based on the collective user response it generates.
  We have identified five distinct categories of retweeting activity on Twitter:  automatic/robotic activity, newsworthy information dissemination, advertising and promotion, campaigns, and parasitic advertisement. 
In the course of our investigations, we have  shown how Twitter can be exploited for promotional and spam-like activities. The content-independent,  entropy-based  activity classification method is computationally efficient, scalable and robust to sampling and missing data. It  has many applications, including  automatic spam-detection, trend identification, trust management,  user-modeling, social search  and content classification on online social media. 
\end{abstract}


\section{Introduction}
\label{sec:intro}
Twitter has emerged as a critical factor in information dissemination, marketing~\cite{WuWatts11}, and influence discovery~\cite{Cha10icwsm}. It has also proven itself to be an important tool for mobilizing people for political protests, as witnessed by the events of the 2011 `Arab spring'~\cite{cbsnews,smh}, and crisis management, when it was used to reconnect Japanese earthquake victims with loved ones and to provide real time information during the unfolding nuclear disaster~\cite{mashable}. In the cultural arena, Twitter has developed into an effective mouthpiece for celebrities \cite{celebrity}, spawning a generation of stars, like Justin Beiber, and starlets  \cite{bieber}.   As a consequence, we have seen the rise of new social marketing strategies and sophisticated automated promotion campaigns. Information dissemination, advertising, propaganda campaigns, bot retweeting and spamming are some of the many diverse activities occurring on Twitter.
Understanding these {activities and their intent} will lead to better tools for trend identification, spam detection, and improve user modeling and content analysis.

Through examples of  retweeting activity, we illustrate  the richness of Twitter dynamics in Section \ref{sec:Dynamics}.
Differentiating between these diverse activities on Twitter and classifying the short posts is a challenging problem.
For example,  a post that is retweeted multiple times by the same user may be categorized as spam. However, if the same message is of interest to and retweeted by many other users, it can be classified as a successful campaign or information dissemination.  Such judgements are difficult to make based solely on content.
The advent of bots and automatic tweeting services have added another dimension of complexity to the already difficult problem. How do we distinguish human activity from programmed or bot activity, campaigns designed to manipulate opinion from those that capture users' interest, and popular from unpopular content?

We propose a novel, quantitative approach to address these questions in Section \ref{sec:approach}. 
We describe an \emph{information-theoretic method} to characterize the dynamics of retweeting activity generated by some content on Twitter. While content- and language-independent, our method is nevertheless able to categorize content into multiple classes based on how Twitter users react to it: it can separate newsworthy stories from those that are not interesting, campaigns that are driven by humans from those driven by bots, successful marketing campaigns from unsuccessful ones. Previous work provided a binary (such as low-quality vs. high quality content ) or tertiary classification of content based on analysis of content and structure~\cite{Agichtein2008Finding} or user response to it~\cite{Crane08}. However, the rich, heterogenous and complex activity on Twitter necessitates the need for a more detailed characterization.

When a user posts or `tweets' a story, he exposes it to other Twitter users.  We focus on tweets that contain URLs and use these URLs as markers to trace the spread of information or content through the Twitter population. When a later tweet includes the same URL as an earlier one, we say that the new post \emph{`retweets'} the content of the original tweet. We do not require the retweet to contain `RT' string nor check that the user follows the author of the original tweet. Our retweets not only include traditional retweets from the original author's followers, but also conversations about the content associated with that URL and independent mentions of it. The collective user response to the tweet or \emph{retweeting activity} varies with the nature of content and users' interest in it, leading to characteristic dynamic patterns. For example, a popular news story will be retweeted by many different users (but only once by each user),  whereas campaigns will get many retweets but from the same small group of users. Some retweets, however, could be automatically generated, and relying purely on frequency of retweets will mislead as to the popularity of content. The temporal signature of automated retweeting is drastically different from collective human response, allowing us to easily differentiate between these.

Given some content (URL), we characterize its retweeting dynamics by two distributions: distribution of the time interval between successive retweets and distribution of distinct users involved in retweeting. We use \emph{entropy} to measure these distributions. We show that these two numeric features capture much of the complexity of user activity in Section \ref{sec:Dataset}.
Using these features to classify activity on Twitter, we have been able to identify several different types of activity, including marketing campaigns, information dissemination,  auto-tweeting and spam. In fact some of the profiles, that we have correctly identified as engaging in spam-like activities have been  eventually suspended by Twitter. Our simple yet powerful approach can easily separate newsworthy content from promotional campaigns, independent of the language of the content, and provides an objective measure of the value of content to people.


\section{Dynamics of Retweeting \\Activity}
\label{sec:Dynamics}
\begin{figure*}[tbh]
\begin{center}
 \begin{tabular}{ccc}
  \includegraphics[height=1.55in]{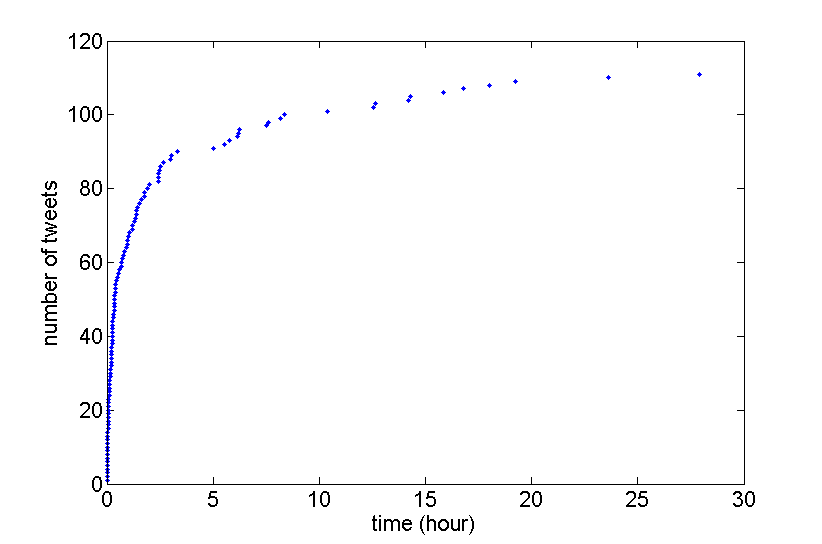}
  &
  \includegraphics[height=1.55in]{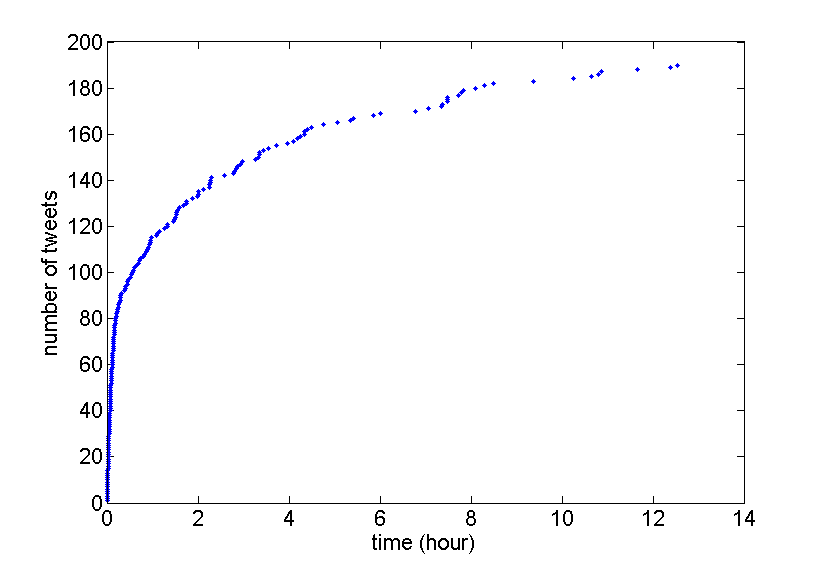}
  &
     \includegraphics[height=1.55in]{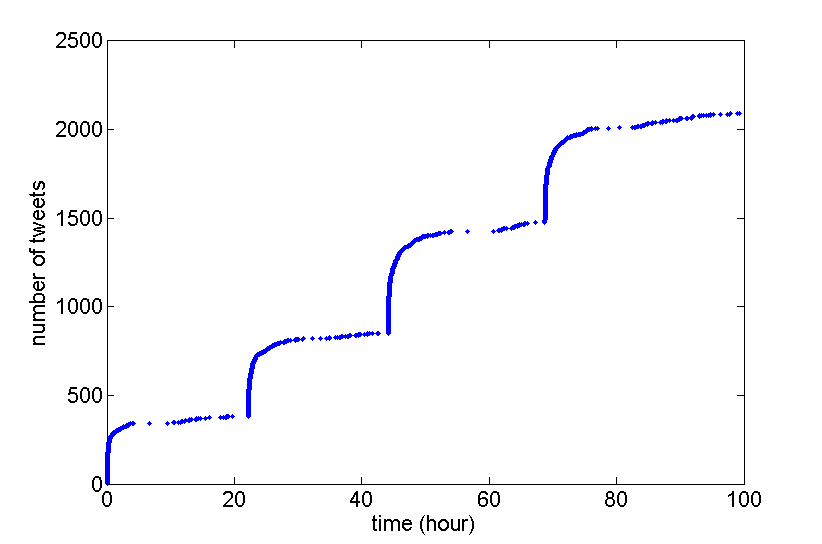} \\
  (a) &   (b)  & (c)\\
     \includegraphics[height=1.55in]{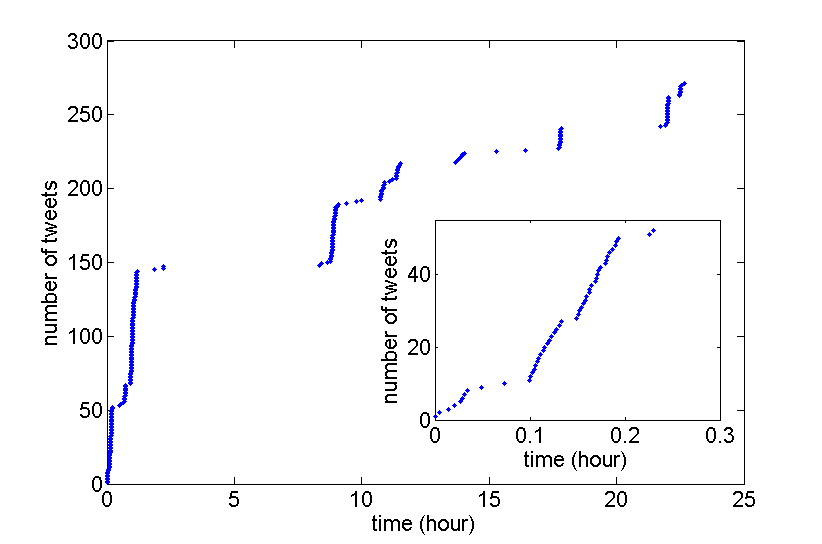}
  &
       \includegraphics[height=1.55in]{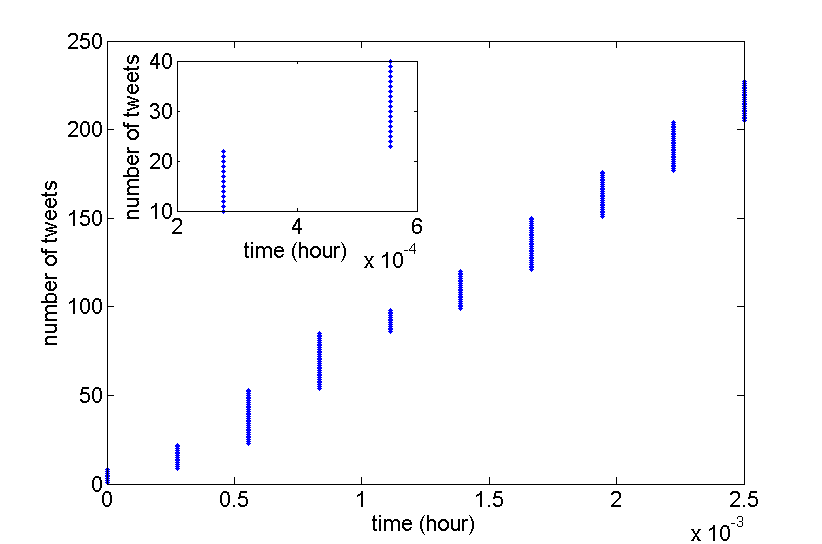}
  &
      \includegraphics[height=1.55in]{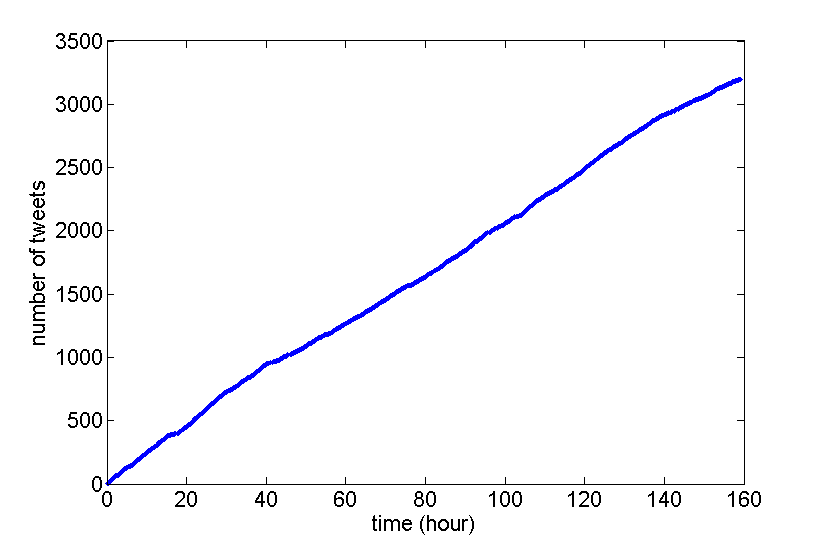}   \\
  (d)  & (e) &  (f)\\
  \includegraphics[height=1.55in]{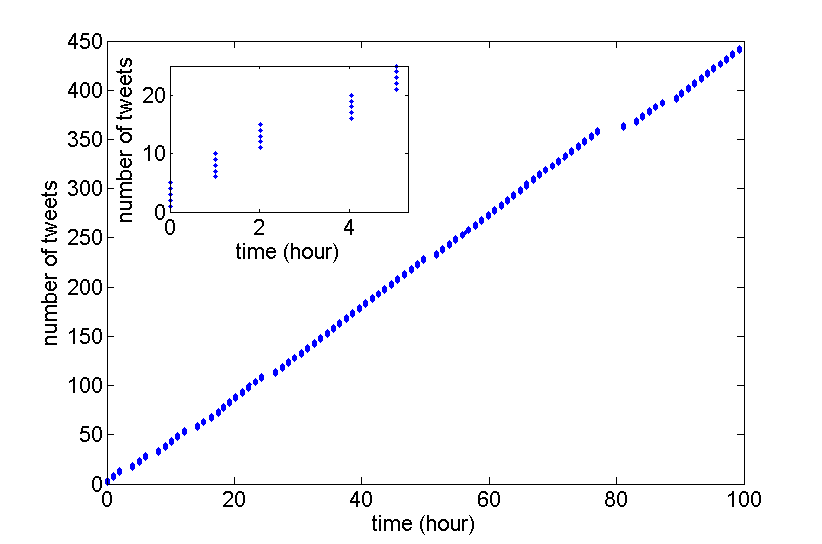}
  &
  \includegraphics[height=1.55in]{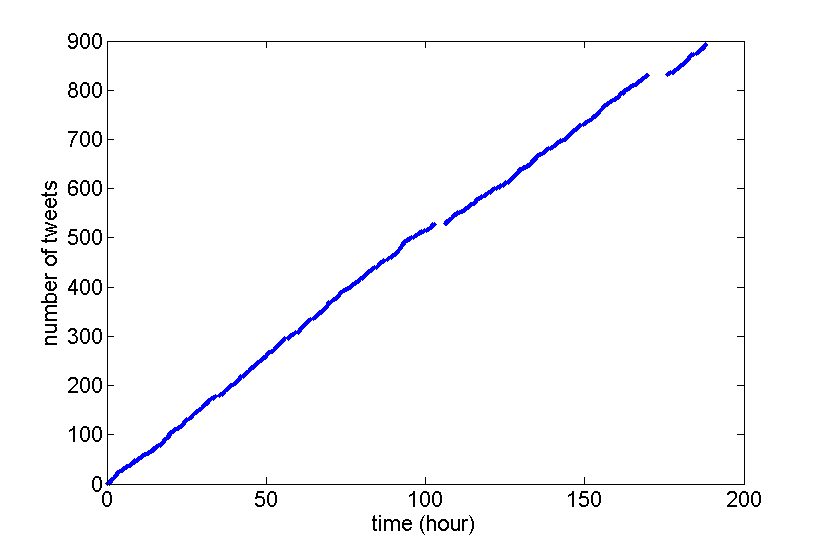}
  &
      \includegraphics[height=1.55in]{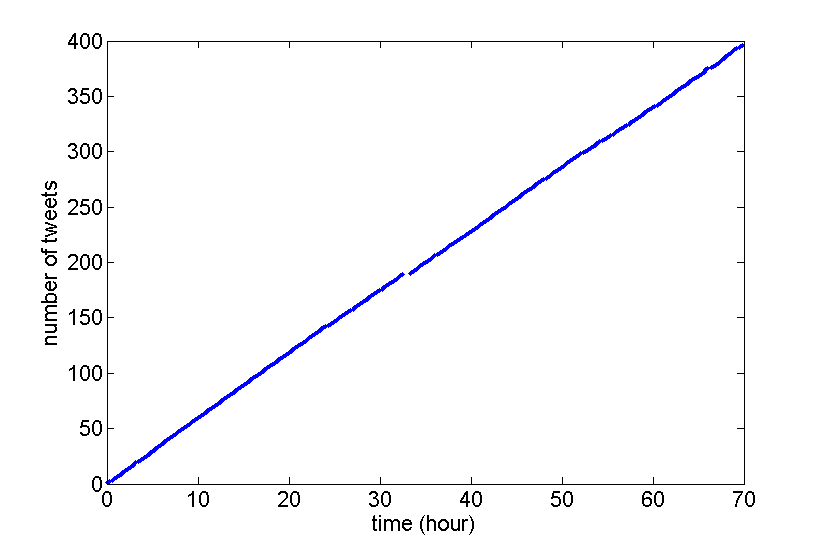}   \\
   (g)  & (h) &  (i)
  \end{tabular}
\end{center}
  \caption{Evolution of retweeting activity for story posted by (a) a popular news website (nytimes) (b) popular celebrity (billgates) (c) politician (silva\_marina) (d) an aspiring artist (youngdizzy)   (e) post by a fan site (AnnieBieber) (f) animal rights campaign (nokillanimalist) (g) advertisement  using social media (onstrategy) (h) advertisement  from an account eventually suspended by Twitter(EasyCash435) (i) advertisement by a Japanese user (nitokono).
Insets in (d), (e) and (g) show automatic retweeting, with multiple retweets made within a short time period either by the same or different users.
}\label{fig:evolution}
\end{figure*}

User's response to content posted on Twitter is encoded in the dynamics of retweeting of this content. Figure~\ref{fig:evolution} shows total number of times nine different URLs were retweeted vs time. The figures show a wide variety of collective user response to content. Figure~\ref{fig:evolution}(a) shows a characteristic response to newsworthy information: fast initial rise followed by a slow saturation in popularity (total number of retweets). Such response is typical  of newsworthy content and information dissemination in online social networks~\cite{Lerman07ic,Lerman10icwsm,Wu07}.  A similar trend is also observed in the response to content (often photos) posted by major celebrities, as Fig~\ref{fig:evolution}(b).

Retweeting activity of posts made by starlets (without major following) is starkly different from that of stars. Figure~\ref{fig:evolution}(d) shows retweeting activity of a post by Young Dizzy, an aspiring artist and songwriter. Short bursts of intense activity are followed by long periods of inactivity. As we show later, this is one of the characteristics of automated tweeting, an increasingly popular feature on social media. In many of these cases, such automated retweets are generated by one or a small groups of users, pointing to attempts to manipulate the apparent popularity of content. Such automated methods to boost popularity are used not only by aspiring starlets, but also by dedicated fans of major stars, e.g., Justin Beiber as shown in Figure~\ref{fig:evolution}(e). In this case, fans are asked to register their Twitter accounts on a website, which then automatically tweets posts about the star from their accounts. There are other example where users (or a small group of users) retweet the same message multiple times, often with the aid of some automated service, leading to a spam-like advertisement campaign. This is shown figures Figure~\ref{fig:evolution}(g) and Figure~\ref{fig:evolution}(h). One of these accounts EasyCash435 was eventually suspended by Twitter. Figure \ref{fig:evolution}(i) shows similar characteristics of some content in Japanese. Note, that using only the retweet dynamics, without any knowledge of the content, we are able to deduce the spam-like advertisement campaign that this profile engages in.  This is confirmed by analysing content.  

In addition to information dissemination, automated tweeting, promotional activities and advertisements, campaigns  add to the diversity of Twitter dynamics. One of the successful campaigners in our sample was a Brazilian politician Marina Silva. Figure~\ref{fig:evolution}(c) traces the retweeting activity of a post made by her over a period of 4 days. Every day she posts the same link using the social media dashboard Hootsuite ({www.hootsuite.com}). The retweeting activity follows a news-like trace seen in (a) and (b). However, when the activity gradually slows down,  she breathes new life into  the campaign by retweeting the same URL, generating a new upsurge in interest (and retweeting).
Contrast this with an not-so-popular animal rights campaign  shown in Figure~\ref{fig:evolution}(f), where the same few users (as shown later) are repeatedly manually retweeting some content to raise its visibility.


\section{Entropy-Based Analysis}
\label{sec:approach}

Manual analysis of retweeting activity on Twitter is labor-intensive. Instead,  in this section we describe a principled approach to categorize retweeting activity associated with some content.

\paragraph{Problem Statement} Given some user-generated content or tweet $c_j  \in C$(where $C$ is a set of tweets or content), our aim is to analyze the trace, $\mathcal{T}_j  \in \mathcal{T} $ (where $\mathcal{T}$  is the collective activity on all content), of retweeting activity on it,  to understand the content and associated dynamics. This trace, $\mathcal{T}_j$ can be represented by a sequence of tuples $((u_{j1},t_{j1}), $$(u_{j2},t_{j2}), $ $\cdots, $ $(u_{ji},t_{ji}),$ $ \cdots$, $(u_{jK},t_{jK}) )$, where $u_{ji}$ represents a user retweeting $c_j$ at time $t_{ji}$. Given $N$ such  traces $\mathcal{T}_1,\cdots , \mathcal{T}_N  \in \mathcal{T}$  and their corresponding tweets $c_1,\cdots, c_j, \cdots, c_N \in C$  , how do we meaningfully characterize and categorize them?

\subsection{Time Interval Distribution}
\label{sec:distribution}

\begin{figure*}[tbh]
\begin{center}
 \begin{tabular}{ccc}
  \includegraphics[height=1.55in]{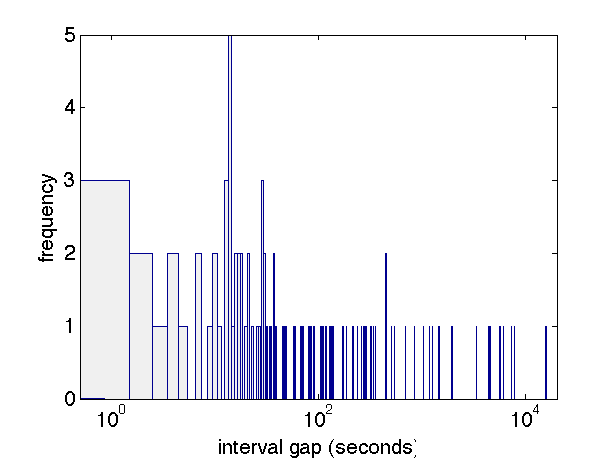}
  &
  \includegraphics[height=1.55in]{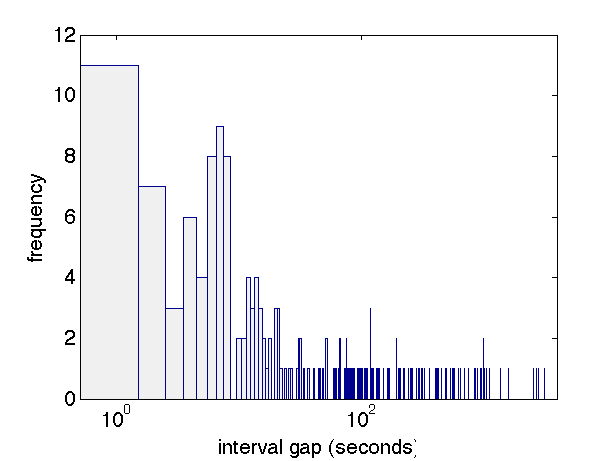}
  &
     \includegraphics[height=1.55in]{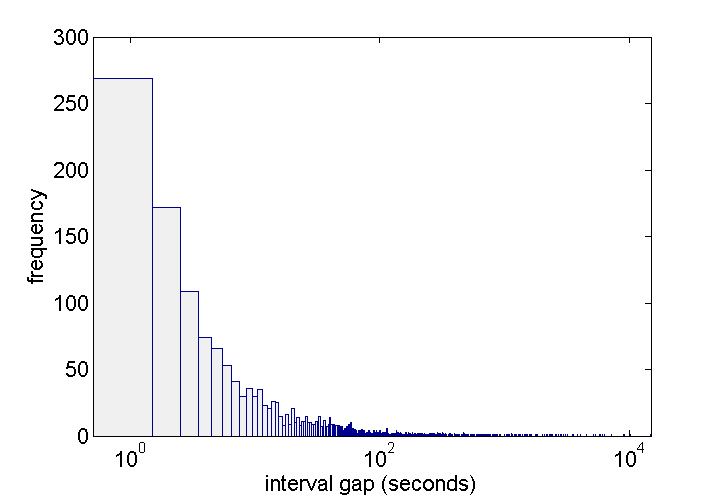}
  \\
  (a) &    (b)  & (c)\\
     \includegraphics[height=1.55in]{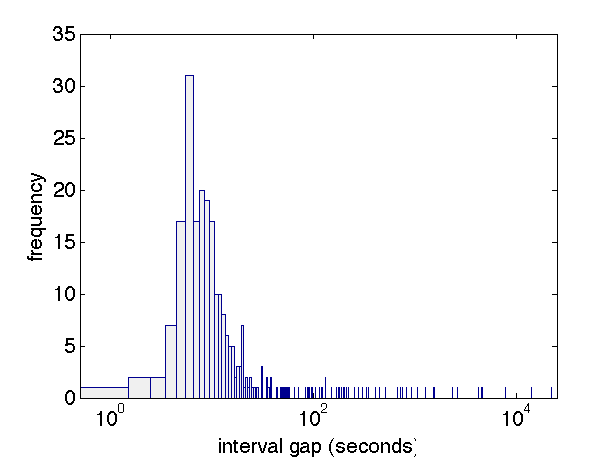}
  &
     \includegraphics[height=1.55in]{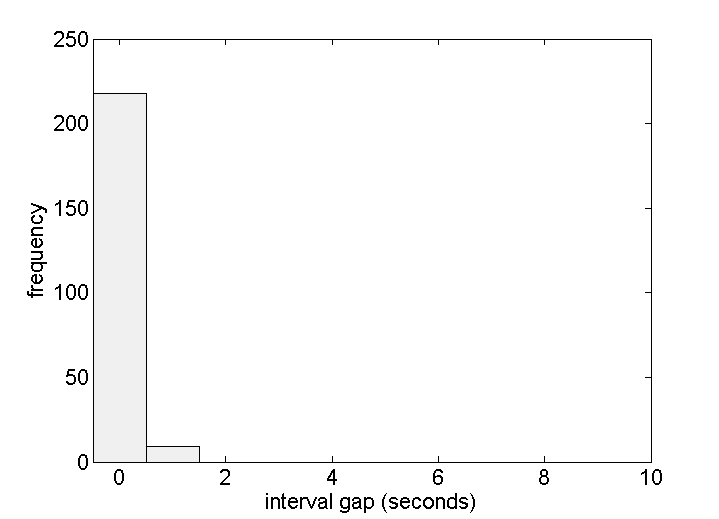}
  &
    \includegraphics[height=1.55in]{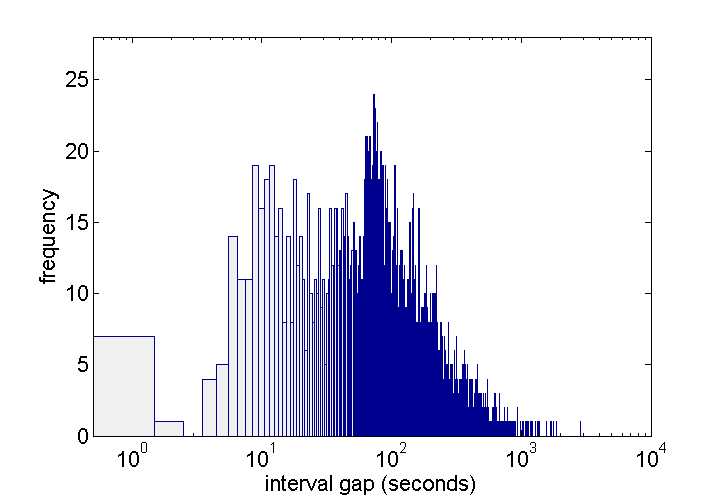}
  \\
  (d)  & (e) &   (f)\\
    \includegraphics[height=1.55 in]{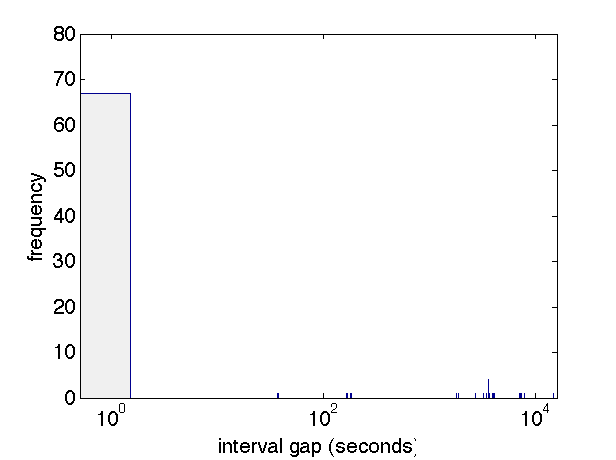}
    &
      \includegraphics[height=1.55in]{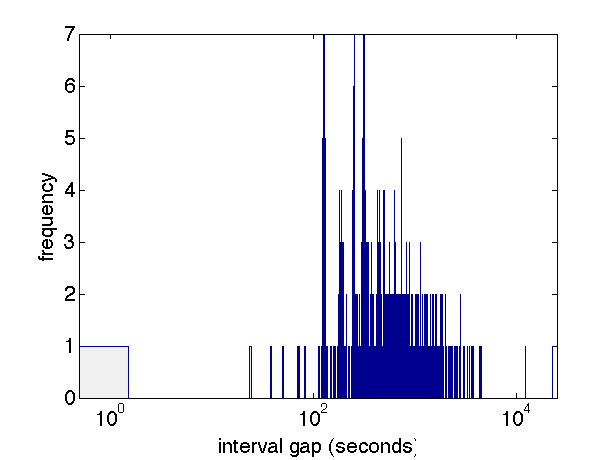}
   &
      \includegraphics[height=1.55 in]{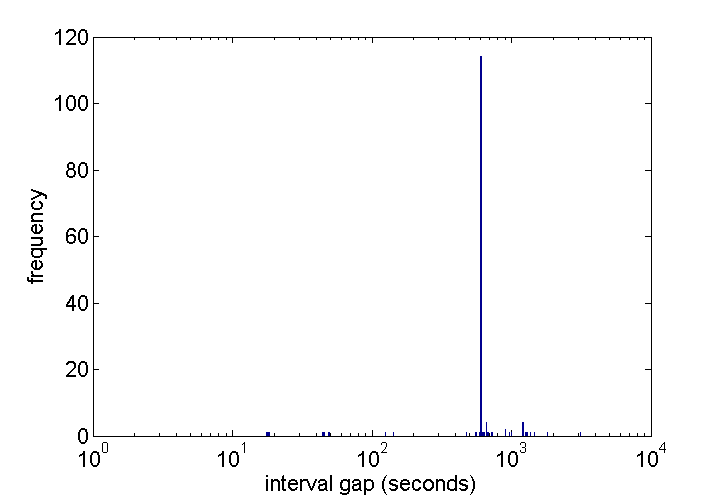}
  \\
   (g)  & (h) &   (i)
  \end{tabular}
\end{center}
  \caption{ The distribution of the inter-arrival gaps for the retweeting activities shown in Figure \ref{fig:evolution} (a)nytimes (b)billgates (c) silva\_marina (d) youngdizzy1 (e) AnnieBieber (f) nokillanimalist  (g) onstrategy (h) EasyCash435 (i) nitokono.  } \label{fig:temporal}
\end{figure*}

The observations we made above about dynamics of retweeting can be succinctly captured by two distributions: inter-tweet time interval  distribution and user distribution. First we consider the distribution of time intervals between successive retweets. These are shown in Figure~\ref{fig:temporal} for the same URLs whose retweeting activity is shown in Figure~\ref{fig:evolution}. 
Humans are very heterogeneous; therefore, a signature of human activity is a broad distribution with time intervals of many different length that are all equally likely, as shown in Figure~\ref{fig:temporal}(a)-(c) and (f). Specifically, there is a lot of activity initially associated with newsworthy content, which gradually decreases with time, resulting in many short intervals and some long ones, as shown in Figure~\ref{fig:temporal}(a)--(b).  Automated retweeting results in tweets at regular time intervals, which will lead to an isolated peak or peaks in the distribution (as in Figure~\ref{fig:temporal}(i)), or bursty behavior with many zero second intervals (as seen in Figure~\ref{fig:temporal}(e) and (g)).


We measure the regularity  or predictability of the temporal trace of tweets using \emph{time-interval entropy}.
Let ${\Delta T}$ represent the time interval between two consecutive retweets in a trace $\mathcal{T}_j$, with possible values $\{\Delta t_1, \Delta t_2,\cdots, \Delta t_i,\cdots, \Delta t_{n_T}\}$. If there are $n_{\Delta t_i}$ time intervals of length ${\Delta t_i}$, then $p_{\Delta T}(\Delta t_i)$ denotes the probability of observing a time interval $\Delta t_i$:
\begin{equation}
 p_{\Delta T}(\Delta t_i) = \frac{n_{\Delta t_i}}{\sum_{k=1}^{n_T} n_{\Delta t_k}}
\end{equation}
The entropy $H_{\Delta T}$ of the distribution of time intervals is:
\begin{equation}
H_{\Delta T}(\mathcal{T}_j)=-\sum_{i=1}^{n_T} p_{\Delta T}(\Delta t_i)log(p_{\Delta T}( \Delta t_i))
\label{eq:1}
\end{equation}
Automatic retweeting with a regular pattern has a lower time interval entropy, and is therefore, more predictable, than human retweeting, which is more broadly distributed and less predictable.

\subsection{User Distribution}
\label{sec:distribution}

\begin{figure*}[tbh]
\begin{center}
 \begin{tabular}{ccc}
  \includegraphics[height=1.55in]{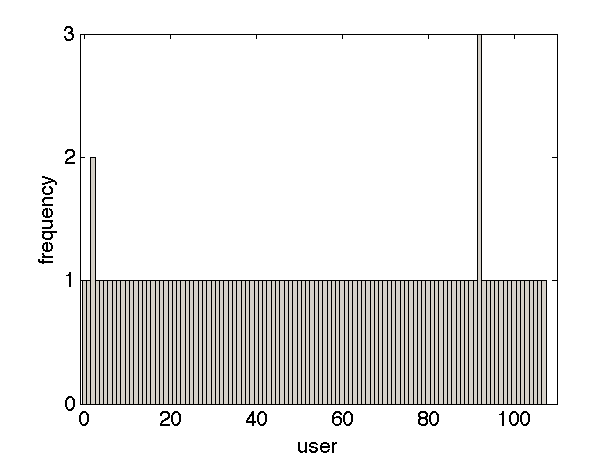}
  &
  \includegraphics[height=1.55in]{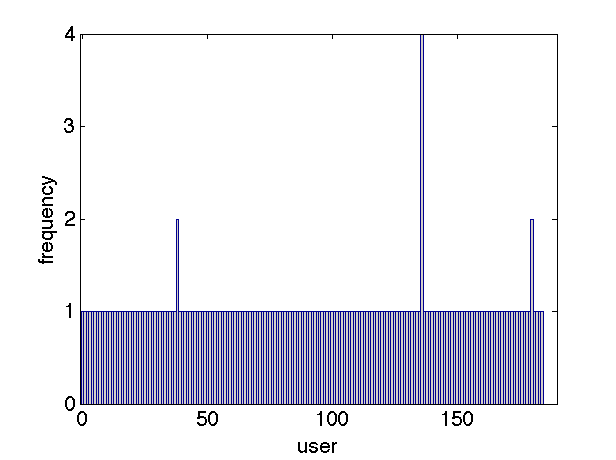}
  &
     \includegraphics[height=1.55in]{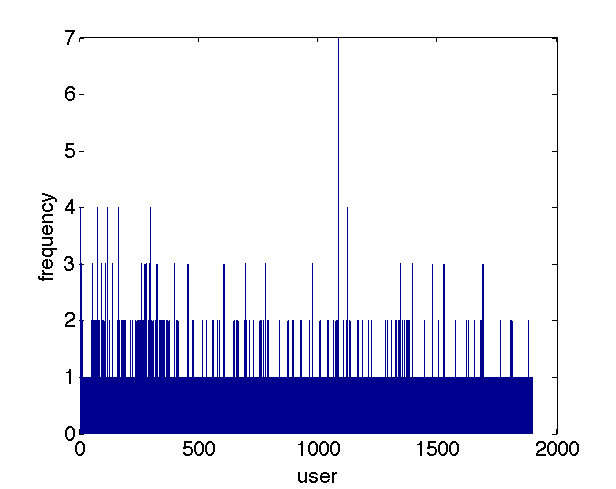}   \\
  (a) &    (b)  & (c)\\
     \includegraphics[height=1.55in]{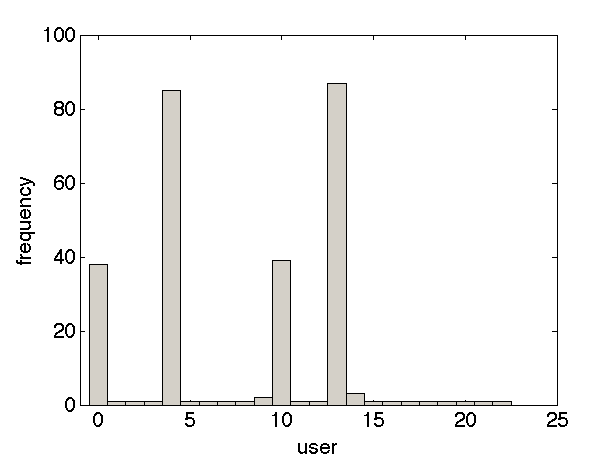}
  &
     \includegraphics[height=1.55in]{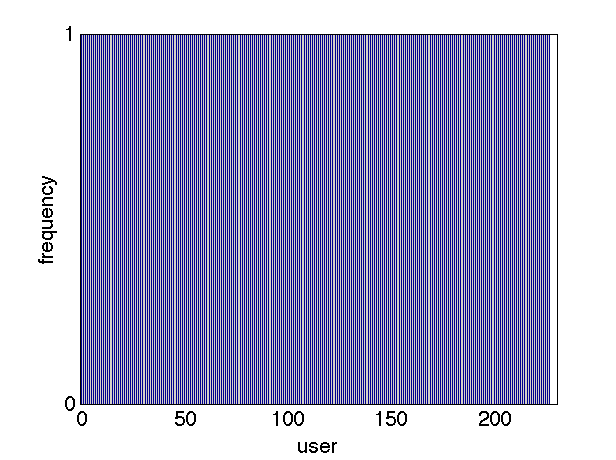}
  &
    \includegraphics[height=1.55in]{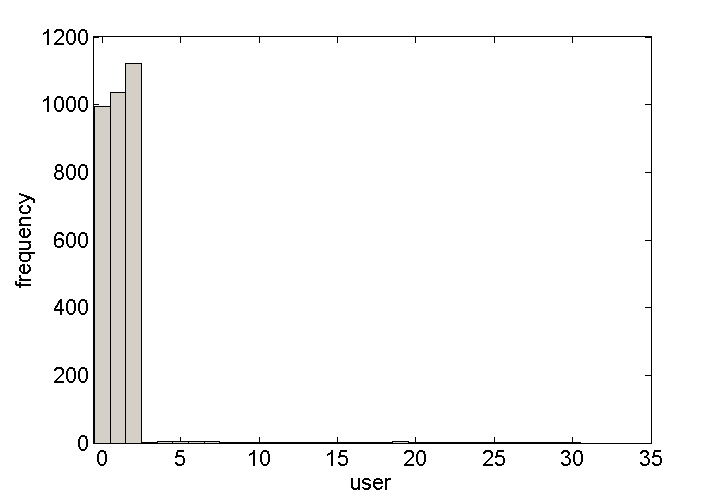}  \\
  (d)  & (e) &  (f)\\
    \includegraphics[height=1.55 in]{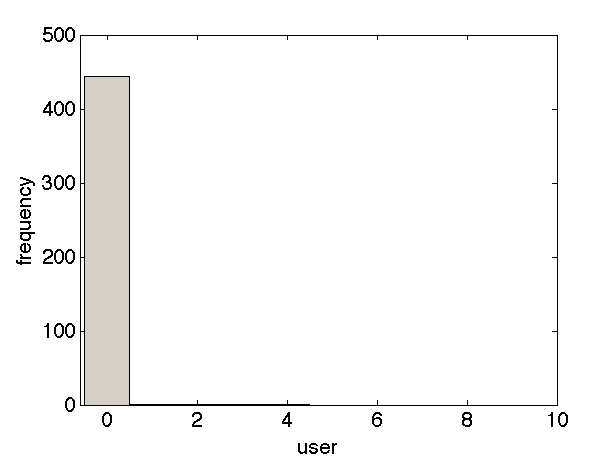}
    &
      \includegraphics[height=1.55in]{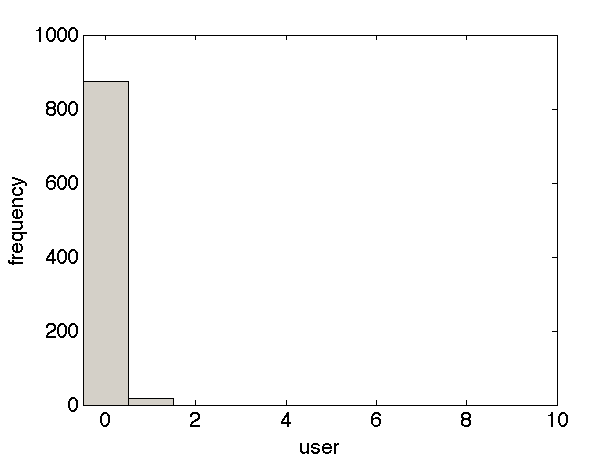}
   &
      \includegraphics[height=1.55 in]{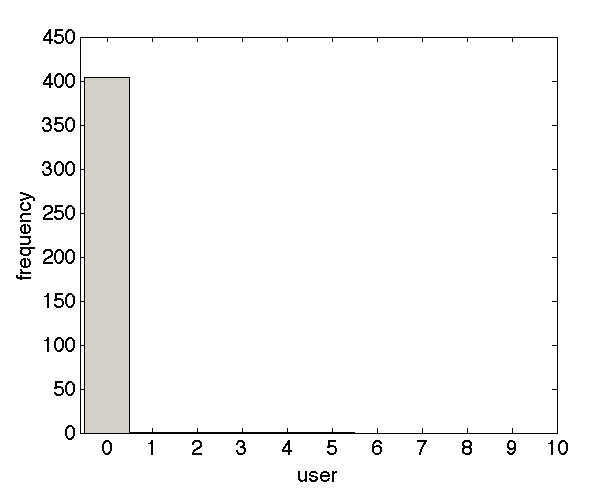}  \\
   (g)  & (h) &  (i)
  \end{tabular}
\end{center}
  \caption{  The number of retweets by distinct users. Each user is marked by a unique user id for the retweeting activities shown in Figure \ref{fig:evolution} (a)nytimes (b)billgates (c) silva\_marina (d) youngdizzy1 (e) AnnieBieber(f) nokillanimalist  (g) onstrategy (h) EasyCash435 (i) nitokono.  } \label{fig:frequency}
\end{figure*}

In addition to time interval, we also measure the distribution of the number of times a specific user retweets some URL. Figure~\ref{fig:frequency} shows the number of retweets made by each unique user involved in the tweeting activity shown in Figure~\ref{fig:evolution}. Newsworthy content is usually retweeted once by each user who participates in the tweeting activity, as shown in Figure~\ref{fig:frequency}(a)--(c).  Spam-like activity and campaigns, on the other hand, result when an individual (Figure~\ref{fig:frequency}(g)--(i)) or a small group (Figure~\ref{fig:frequency}(f)) repeatedly retweet the same post. The higher the retweeting, the greater the manipulation effort. 

The campaign shown in Figure~\ref{fig:evolution}(c) is successful, since there are many distinct users who participate in it, as shown in Figure~\ref{fig:frequency}(c). However, there are some dedicated campaigners, including $silva\_marina$ herself, who retweet the same message multiple times. Also the distribution of inter-arrival times in Figure~\ref{fig:temporal}(c) is similar to that of Figure~\ref{fig:temporal} (a) and (b), indicating human activity. A campaign probably not as successful as that by $silva\_marina$ is one by $nokillanimalist$  (Figure~\ref{fig:evolution}(f)), which has very few participating users in it. The distribution  of the inter-arrival times in Figure~\ref{fig:temporal}(f) is also comparable to Figure~\ref{fig:temporal}(a)--(c), with a large number of nonzero inter-arrival times and the frequency of shorter inter-arrival gaps being larger than that longer ones indicating human activity. However, the distribution of the number of retweets by distinct users shows a stark contrast.  In fact it shows that there are only three dedicated users generating over 3000 retweets.  Similarly in case of of the retweeting activity shown in Figure~\ref{fig:evolution}(h), there are only two users engaged in spreading spam-like advertisements (Figure \ref{fig:frequency}(h)). These two users together account for around 900 retweets. Spam-like characteristics are also observed in the advertisements, whose retweeting activity is shown in Figure \ref{fig:evolution}(g) and \ref{fig:evolution}(i) which have one (Figure \ref{fig:frequency}(g)) and two users (Figure \ref{fig:frequency}(i)) generating a bulk of the content. However on looking into the temporal distribution more closely, we observe that in case of Figure \ref{fig:evolution}(g),
 almost two-thirds of the retweets occur almost consecutively (time interval gap is zero seconds), indicating a possible autotweeting activity. 
Figure \ref{fig:evolution}(i) too, shows some kind of  probable scheduled or automated tweeting activity with around 28\% of the tweets having an exact interval gap of 604 seconds. Possible autotweeting is also indicated in the promotional activity shown in Figure \ref{fig:evolution}(e). Although a large number of users participate in this activity as shown by Figure \ref{fig:frequency}(e), almost all the retweets are generated simultaneously as seen in Figure \ref{fig:temporal}(e).

We also use entropy to measure the breadth of user distribution. Let random variable $F$ represent a  distinct user in a trace $\mathcal{T}_j$, with possible values $\{f_1,f_2,\cdots,f_i,\cdots,f_{n_F}\}$. Let there be $n_{f_i}$ retweets from user  ${f_i} $ in the trace $\mathcal{T}_j$.
If $p_F$ denotes the probability mass function of $F$, such that $p_F(f_i)$ gives the probability of a retweet  being generated by user  $f_i$, then
\begin{equation}
 p_{F}(f_i) = \frac{n_{f_i}}{\sum_{k=1}^{n_F} n_{f_k}}
\end{equation}
The user entropy $H_F$ is given by:
\begin{equation}
H_F(\mathcal{T}_j)=-\sum_{i=1}^{n_F} p_{F}(f_i)log(p_{F}(f_i))
\label{eq:2}
\end{equation}

As clear from the Equation \ref{eq:2}, spam-like activity  having small number of users responsible for large number of tweets, would have lower entropy than  retweeting activity of newsworthy content. On the other hand, automated retweeting coming from many distinct users (as in Figure~\ref{fig:frequency}(e)) would indicated that users' accounts may have been compromised.




\subsection{Classification}
Time interval and user entropies $H_{\Delta T}(\mathcal{T}_j)$ and $H_F(\mathcal{T}_j)$) can used to categorize retweeting activity of the given content. This classification helps us not only identify the different dynamic activities occurring on Twitter, but also, provides us valuable insight on the nature of the associated content.

The linear runtime complexity of entropy calculation and the presence of scalable methods of clustering \cite{Bradley2000Clustering} ensures that this entropy-based approach can easily be applied to very large datasets.

\section{Validation}
\label{sec:Dataset}

Twitter's Gardenhose streaming API provides access to a portion of real time user activity, roughly 20\%-30\% of all user activity. We used this API to collect tweets for a period of three weeks in the fall of 2010. We focused specifically on tweets that included a URL (usually shortened by a service such as bit.ly) in the body of the message. In order to ensure that we had the complete retweeting history of each URL, we used Twitter's search API to retrieve all activity for that URL. 

Data collection process resulted in 3,424,033 tweets which mentioned 70,343 distinct shortened URLs.  There were 815,614 users in our data sample.
We study the retweeting activity of URLs posted by users who posted at least two popular URLs. By popular, we mean URLs that were retweeted at least 100 times. There were 687 such  distinct URLs.

We apply entropy based approach to study the retweeting dynamics of these URLs. We show that entropy-based analysis gives a good characterization of different types of activities observed in collective retweeting of these URLs.

 \subsection{Manual Annotation}
We manually examined the content of each URL (using Google translate on foreign language pages) to annotate the activity along following categories:
\begin{description}
\item[News] If the URL belongs to the twitter profile of a news organization, we label the retweeting activity as  following news.
\item[Blogs] If the URL links to the blog or webpage maintained by an individual, we classify the retweeting activity as following blogs or celebrity.
\item[Campaigns] If the URL belongs to an individual or an organization with a discernible agenda (politics, animal rights issues), we classify the retweeting activity as campaign.
\item[Advertisements and promotions] If the URL links to an advertisement or promotion, we classify the retweeting activity as such. This includes instances where users post the same link repeatedly, leading to spam-like content generation, and the promotional activities of aspiring starlets.
\item[Parasitic ads]  This is a form of parasitic advertisement in which users participate unwittingly. This happens when a user logs into a website or web service, and then that service tweets a message in user's name telling his followers about it.  For example, when a user visits sites such as Tinychat \footnote{tinychat.com} or Twitcam \footnote{twitcam.com}, a message is posted to the user's Twitter account ``join me on tinychat...''
\item[Automated/robotic activity] Retweeting that is mainly generated through Twitterfeed \footnote{www.twitterfeed.com} or similar services is classifies as automatic activity. Note that automated activity could be associated with any type of content, but since it has its own unique characteristics, different from all the aforementioned activities,  we included it as a separate class. This can be easily identified by looking at the source of the tweet, which will identify $twitterfeed$ (or a similar service) as the originator.
\end{description}
We found that users respond to news stories and blog posts in identical manner, making them difficult to distinguish. Generally, the type of information contained in these two sources is also very similar. Therefore, for classification purposes, we put them in the same category of newsworthy content.

\begin{figure}[tbh]
  \includegraphics[width=\columnwidth]{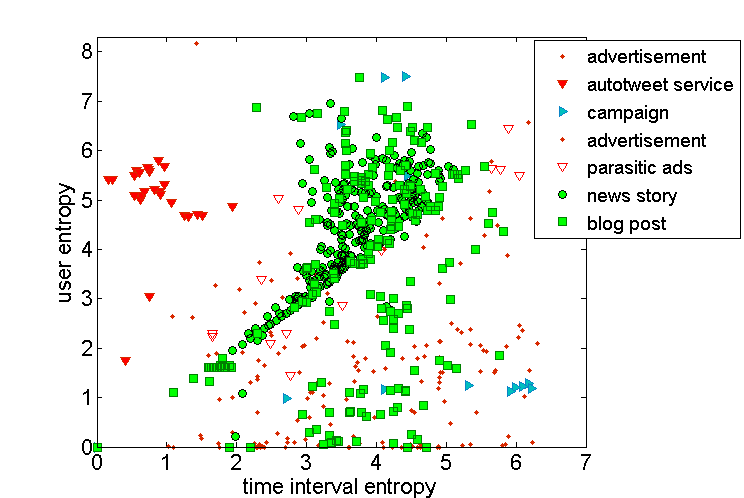}
 \caption{Manually annotated URLs shown in the entropy plane.
} \label{fig:manual}
\end{figure}

Figure~\ref{fig:manual} shows the retweeting activity of URLs in our data sample as measured by the time interval and user entropy. The bulk of the URLs belong to news or blog category. They are also characterized by medium to high user entropy and time interval entropy, indicating newsworthy content. Blog posts or websites of major celebrities represent more popular content and are located in the upper section of the plot. Blog posts from starlets without major following are located in the lower section of the plot. Though these posts have similar numbers of retweets, lower user entropy means that the starlets, or their dedicated followers, generate much of the retweeting activity. The automatic retweeting cluster is isolated. This contains URLs like one whose activity is shown in Figure~\ref{fig:evolution}(e), but also several news stories, most notably from the online technology magazine TechCrunch. This is because many Twitter users employ Twitterfeed to automatically tweet stories that are posted on TechCrunch. This helps users appear to be more active on Twitter than they really are. The uninteresting stories are not retweeted by other people. They have low time interval entropy due to automatic retweeting, but high user entropy, since many different Twitter users are associated with the activity.

Advertisements are mostly located in the lower half of the figure, although successful advertisements that capture public interest are indistinguishable from newsworthy content. Unsuccessful campaigns that are driven by a few dedicated zealots are in their own cluster with high time interval and low user entropy, but successful campaigns are also indistinguishable from newsworthy content.

\subsection{Classification}
\label{sec:entropy}
The distribution of distinct time intervals and users involved in the retweeting activity gives a good characterization of the retweeting activity.  As explained in Section \ref{sec:approach}, temporal and user entropy are used to quantify these distributions. Temporal entropy is maximum when the time intervals between any two successive retweets is different. User entropy is maximum when each user retweets the message only once. Next, using temporal and user entropies as features, we classify the retweeting activity represented by a trace ${T}_j  \in \mathcal{T} $. We perform both unsupervised and supervised classification. The data is manually labelled to  train the supervised classifier and to evaluate the performance of the classification techniques. We used Weka software library \footnote{www.cs.waikato.ac.nz/ml/weka} for off-the-shelf implementation of EM (expectation maximization \cite{dempster77em}),k-NN (k-nearest neighbors) and SVM(support vector machines \cite{BoserGV92}) classification.

 \subsubsection{Supervised Classification}
\begin{table}
   \caption{F-Measure (F) and ROC area for 10-fold cross validation experiments using SVM and k-NN classification  }
    \begin{center}
\begin{tabular}{@{}c@{}c@{}|@{}c@{}|@{}c@{}|@{}c@{}|@{}c@{}|@{}c@{}|}
\cline{3-7}
& & ads \&	&	auto-	&	campaign	&	news &	{parasitic} \\
&& {promotion}    &   tweet      &    &   \& blog	& ads \\ \hline

\multicolumn{1}{|c|}{\multirow{2}{*}{\small{k-NN}}} &
\multicolumn{1}{|c|}{\small{F}} &  0.686 & 0.96 & 0.5 & 0.89& 0.105   \\ \cline{2-7}
\multicolumn{1}{|c|}{}                        &
\multicolumn{1}{|c|}{\small{ROC}}  & 0.807 & 0.959 & 0.678& 0.837 & 0.644  \\ \cline{1-7}
\multicolumn{1}{|c|}{\multirow{2}{*}{\small{SVM}}} &
\multicolumn{1}{|c|}{\small{F}} & 0.719 & 0.939 & 0.526 & 0.897& 0     \\ \cline{2-7}
\multicolumn{1}{|c|}{}                        &
\multicolumn{1}{|c|}{\small{ROC}} &0.833 & 0.973 & 0.685 & 0.875 & 0.718\\ \cline{1-7}
\end{tabular}
\end{center}
\label{tb:3}
\end{table}

We used Support Vector Machine with radial basis function (RBF)  kernel and k-NN algorithm with three nearest neighbors and Euclidean distance function to classify the data. Table~\ref{tb:3} reports results of 10-fold cross validation in each model was trained on 90\% of the labeled data and tested on the remaining 10\%.
The F-scores of both algorithms are relatively high, showing that they have well separated instances into different classes.

  \subsubsection{Unsupervised Classification}
\begin{figure*}[tbh]
\begin{center}
 \begin{tabular}{cc}
  \includegraphics[width=\columnwidth]{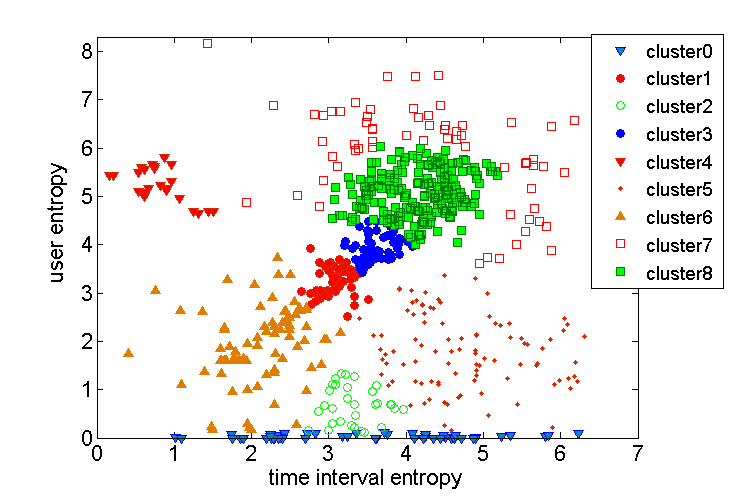} & 
    \includegraphics[width=\columnwidth]{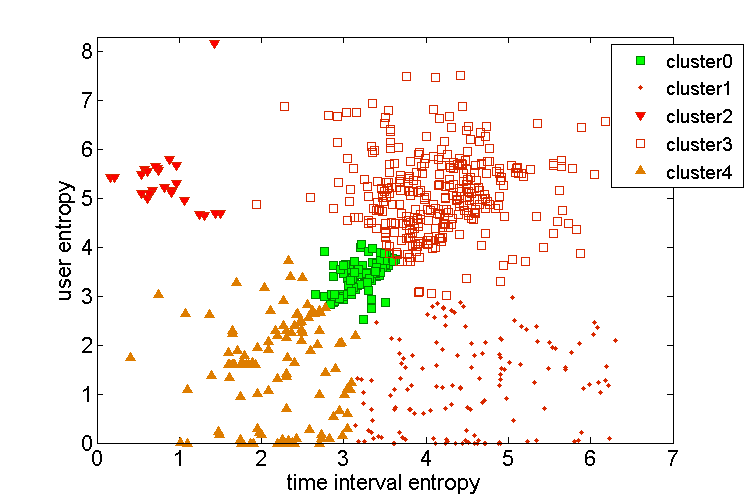} \\
(a) & (b)
\end{tabular}
\end{center}
 \caption{Unsupervised clustering of the data points using EM: (a) when EM automatically finds the best number of clusters, and (b) when the number of clusters is constrained by be five.
} \label{fig:EM}
\end{figure*}

We use Expectation Maximization (EM) algorithm to automatically cluster points. EM uses Gaussian mixture model and can decide how many clusters to create by cross validation. The number of clusters determined automatically by this method was nine.  Figure~\ref{fig:EM}(a) shows the resulting clusters, and the confusion matrix is shown in Table \ref{tb:2}. If the number of clusters were predefined to be 5, the resulting confusion matrix is shown in Table~\ref{tb:1}, and discovered clusters are shown in Figure~\ref{fig:EM}(b).

 \begin{table*}
 \caption{Confusion matrix with manually annotated data and clusters automatically detected by EM algorithm}
 \begin{center}
\begin{tabular}{|c|c|c|c|c|c|c|}
\hline
	&	advertisement	&	auto-tweet	&	campaign	&	news & blogs	&	parasitic \\
    & \& promotion      &               &               &        &          & advertisement \\
	\hline	
cluster0	&	45	&	0	&	0	&	0	&	8	&	0	\\	\hline
cluster1	&	7	&	0	&	0	&	41	&	13	&	1	\\	\hline
cluster2	&	17	&	0	&	0	&	0	&	14	&	0	\\	\hline
cluster3	&	0	&	0	&	0	&	53	&	10	&	1	\\	\hline
cluster4	&	0	&	23	&	0	&	0	&	0	&	0	\\	\hline
cluster5	&	53	&	0	&	7	&	2	&	34	&	0	\\	\hline
cluster6	&	36	&	2	&	1	&	27	&	19	&	6	\\	\hline
cluster7	&	10	&	1	&	3	&	14	&	30	&	6	\\	\hline
cluster8	&	11	&	0	&	2	&	130	&	60	&	0	\\	\hline
\end{tabular}
\end{center}
\label{tb:2}
\end{table*}

\begin{table*}
  \caption{Confusion matrix with manually annotated data and clusters  detected by EM algorithm when number of clusters is predefined to be 5}
 \begin{center}
\begin{tabular}{|c|c|c|c|c|c|}
\hline
	&	advertisement	&	auto-tweet	&	campaign	&	news &	parasitic \\
    & \& promotion      &               &               &   \& blogs	 & advertisements\\
\hline	
cluster0	&	7	&	0	&	0	&	82	&	1\\
\hline
cluster1	&	85	&	0	&	7	&	49	&	0 \\
\hline
cluster2	&	1	&	23	&	0	&	0	&	0\\
\hline
cluster3	&	22	&	1	&	5	&	272	&	7\\
\hline
cluster4	&	64	&	2	&	1	&	52	&	6 \\
\hline
\end{tabular}
\end{center}
\label{tb:1}
\end{table*}

    \subsubsection{Observations}
    \label{sec:results}
    Broadly speaking, we identify five classes of retweeting activity and associated content on Twitter.
      \paragraph{Automatic/Robotic Activity}
        As we can see from the results, almost all methods classify automatic or robotic retweeting (auto-tweet) with high accuracy. While some of such activity, in our data set is related to technology news stories, and their user entropy is similar to that of other news stories, such activity has a much lower time interval entropy than other news stories. 

Two primary kind of automated services that we identified are auto-tweeting services and tweet-scheduling services.
There are two categories of auto-tweeting activities.
The first arises when an individual subscribes to an automatic service that tweets messages on the user's profile on his behalf. One such automatic service is Twitterfeed \footnote{www.twitterfeed.com}, through which the user can subscribe to a blog or news website (any service with an RSS feed). Twitter users employ this service to automatically retweet stories posted on technology news sites Mashable and TechCrunch. This leads to \emph{individual auto-tweets} observed from the profile of that user. 

However, this auto-tweeting feature is also being used for promotional and perhaps phishing activities. For example, a fan site ({http://bieberinsanityblog.blogspot.com/}) for Justin Bieber asks fans to provide their Twitter account information.  The site is  powered by Twitterfeed, and then auto-tweets Justin Bieber news from the profiles of registered fans, resulting in \emph{collective auto-tweeting}. 

Services like  Tweet-u-later \footnote{http://www.tweet-u-later.com/} and Hootsuite can be used to schedule tweeting activities. These websites can be used for spamming. Registering a collection of profiles to these websites and scheduling the a tweet to posted repeatedly, enables spammers to post the same message multiple times.

   Since our method can easily differentiate human activity from bot or automated activity, we are able to identify marketing companies which engage automated services to increase their visibility on Twitter. Such services include OperationWeb ({http://www.operationweb.com/}) and TweetMaster ({http://tweetmaster.tk/}), which claim that they ``will tweet your ad or message on my Twitter accounts that add up to over 170k* followers 2-6 times per day for 30 days.''

Most of these services use bots or automated services to push up the perceived visibility of  the advertisements. To increase visibility they need a large number of profiles. To gain access to a large number of profiles, such services ask users to register, set their own prices for tweets and feature the sponsored tweets in their profile. In this way these services create a win-win situation, helping companies to promote their product and users to make money by featuring sponsored messages on their profiles.

    \paragraph{Newsworthy information}
  This class comprises of mostly news and blogs and some successful campaigns.  Newsworthy information is characterized  by comparable (usually high) user and temporal entropy. Since people, not bots, are involved in disseminating such content, we call this ``human response to information.'' Both supervised and unsupervised clustering algorithms able to separate news and blogs i.e information sharing by humans, from the rest of retweeting activity with good accuracy (Tables \ref{tb:3}, \ref{tb:1} and \ref{tb:2}). However, EM algorithm with five classes breaks this class into smaller clusters (cluster0, cluster3 and cluster4). This is indeed a meaningful subdivision based on popularity, with content in cluster3 being the most popular, content in cluster0 being normal content and content in cluster4 having low popularity.  When EM is allowed to automatically adjust the number of clusters, the popular clusters found by the earlier algorithm gets subdivided into two more classes giving five clusters of human response to information (cluster1, cluster3, cluster6, cluster7 and cluster8  in Figure~\ref{fig:EM}(b)). Compared to hand-labeled dataset (Figure~\ref{fig:manual}) and from the confusion matrix in Table \ref{tb:2}, we observe that cluster7 comprises predominantly of popular blogs, cluster8 comprises mostly of popular news, cluster1 and cluster3 comprise of normal human response to information and cluster6 shows human response to unpopular information.

\paragraph{Advertisements and Promotions}
Advertisements and promotions are distinguished by low user entropy and low to high temporal entropy.
Supervised clustering is able to  accurately detect advertisements and promotions (Table~\ref{tb:3}). Most spam-like advertisements fall in this section. These are unwanted advertisements which are never retweeted by any user besides the originator of the advertisement. EM algorithm with five classes also identifies a group comprising predominantly of advertisements. However, EM algorithm with automatic class detection, divides this group further into three classes: cluster0 comprising mostly of spam-like activity with very low user entropy ($\approx$ 0), cluster2 containing advertisements with low user and medium time entropy  and cluster5 comprising of campaign-like promotions and advertisements with low user entropy and medium to high temporal entropy.

\paragraph{Campaigns}
 Campaigns are identified by low user entropy and very high temporal entropy. There are very few campaigns in the hand-labeled dataset. Even then,  supervised algorithms are able to classify campaigns with a fair degree of accuracy (cf. Table \ref{tb:3}). However, unsupervised algorithm  merges campaigns with advertisements and promotions. Due to considerable overlap of characteristics of campaigns with advertisements or promotions, to distinguish a campaign from an advertisement is difficult, even for manual annotators.
 Note, that when a campaign is very successful like the one by $silva\_marina$, Figure~\ref{fig:evolution} (c), information that the campaigner intends to propagate spreads through the online social media. The retweeting activity in this case becomes similar to human response to information.

\paragraph{Parasitic Advertisements}
None of the methods were able to identify parasitic advertisements very accurately. One possible reason may be their parasitic nature, where they do not have a distinct characteristic feature of their own, but adopt the characteristics of the hosting user profile.

\section{Related Work}
\label{sec:rel}
There has been some work to define temporal variation on online social media. In  \cite{Yang2011Patterns}, the authors enumerate the different approximate shapes of temporal distribution of  content in Twitter. But unlike us, they are not able to associate semantic meaning to the clusters they observe.

Previous work has tried to estimate the quality or interestingness of content \cite{Agichtein2008Finding,Crane08}. However, quality or interestingness is a subjective measure and is biased by the perspective of the user. For instance, what would be high quality information or interesting to a campaigner might be junk to a news aggregator. Therefore  there is the need for an objective  quantitative measure of user-generated content. Our entropy-based approach for classifying user activity and content addresses this need. While the method described in \cite{Crane08} is similar in spirit to ours, it can discover only three classes of activity. Heterogeneous activity on Twitter requires more than three classes.

Most of the existing  spam detection \cite{Markines09} and trust management systems  \cite{Caverlee2008Socialtrust} are based on content and structure but do not look at collective dynamics. Besides, they usually  require additional constraints like labelled up-to-date annotation of resources, access to content and cooperation   of search engine. Satisfiability of so many constraints is difficult especially when one takes the diversity and astronomical size of online social media into account. Our method on the other hand, while having no such constraints,  may be able to detect spams with an accuracy close to humans.

There has been some work done on spam detection on Twitter.
Grier et. al \cite{grier:spam} analyzed the features of spam on Twitter.  However, they detect spam using three blacklisting services.  Similarly,  one of the methods employed to remove spam on Twitter is using Clean Tweets\footnote{http://www.seoq.com/blvdstatus/clean-tweets.html}~\cite{Kwak2010What}. Clean tweets filter tweets from users who are less than a day (or any duration specified)  old and tweets that mention three (or any number specified) trending topics. However, it would be unable to detect spammers  who auto-tweet or posts spam-like tweets at regular intervals (like EasyCash435 or onstrategy,  Figure \ref{fig:evolution} (g) and (h)), which our approach can easily detect. Also, since URL shortening services such as {http://bit.ly} are often used on Twitter, users cannot guess which references are pointed at, which in turn is an attractive feature for spammers. However, since our categorization method is content independent, we can easily identify such spams using this method.
Yardi et al.~\shortcite{yardi:spam}  state ``Twitter spam varies in style and tone; some approaches are well-worn and transparent and others are deceptively sophisticated and adaptable.''  Using this method, we can capture the characteristics of spam, whether, it is generated by a auto-tweet service, a malicious advertiser or a passionate campaigner.

Automated email spamming has been studied by \cite{Xie08}. They have identified the activity of botnets generating e-mail spam as being `bursty' (inferred from the duration of activity) and `specific' (pertaining to a random generated URL matching the signature). In  this study of Twitter, we identify automated activity by a set pattern of retweeting (indicated by much lower time-interval entropy compared to  user entropy).

Note, that this approach is based on observed  collective response to content. By reposting  some content, users give an implicit feedback to that content. Therefore unlike\cite{Agichtein2008Finding, Caverlee2008Socialtrust}  this method can even be applied in systems, where users do not explicitly rate other users.

We can automatically detect newsworthy, information-rich content and separate it from other user-generated content, based on user-response. We have showed that this method can further categorize content within this class into blogs or celebrity websites and news. \cite{WuWatts11} study the flow of information between these sub-categories.


\section{Conclusion and Future Work}
\label{sec:conclusion}
We characterize dynamics of retweeting activity associated with some content on Twitter by the entropy of the user and time interval distributions, and show that these two features alone are able to separate user activity into different  meaningful classes. The method is computationally efficient and scalable, content and language independent and is robust to missing data. Entropy-based classification can be used for spam detection, trend identification, trust management, user modeling, understanding intent and detecting suspicious activity on online social media. We have identified five categories of retweeting activity on Twitter: newsworthy information dissemination, advertisements and promotions, campaigns, automatic or robotic activity and parasitic advertisements.   We have observed that human response to news, blogs and celebrity posts is very similar. The novel entropy-based classification method not only enables us to characterize user activity, but it also helps us to understand user-generated content and separate popular content from normal or unpopular content.  Future work includes analyzing the effect of adding more features to the categorization of user activity and content. We also plan to apply this analysis on larger datasets and other online social media.
In the course of our study we have observed the gradual emergence of sophisticated spamming and the birth of an alternate industry to manipulate content on Twitter like promotional activities to improve the perceived popularity of stars. Kwak et al. \cite{Kwak2010What}  had asked an important question -- What is Twitter, a Social Network or a News Media? Our analysis of Twitter shows that it is not only both a social network but much more -- the diversity of twitter activity is a reflection of complexity of collective user dynamics on online social media.

\small
\bibliographystyle{abbrv}
\bibliography{krisl-for-snakdd11,rumig-for_sna_kdd11}  
\balancecolumns
\end{document}